# van der Waals forces control the internal chemical structure of monolayers within $ABP_2X_6$ lamellar materials

Sherif Abdulkader Tawfik,[1,2] Jeffrey R. Reimers,[1,3] Catherine Stampfl,[4] and Michael J. Ford[1]

[1]School of Mathematical and Physical Sciences, University of Technology Sydney, Ultimo, New South Wales 2007, Australia
[2]Institute for Biomedical Materials and Devices (IBMD), Faculty of Science, University of Technology Sydney, Sydney, NSW, Australia
[3]International Centre for Quantum and Molecular Structures and School of Physics, Shanghai University, Shanghai 200444, China
[4]School of Physics, The University of Sydney, New South Wales, 2006, Australia.

* sherif.abbas@uts.edu.au, *Jeffrey.Reimers@uts.edu.au, Mike.Ford@uts.edu.au*

**Abstract:** Following the recent demonstration that van der Waals forces control the ferroelectric ordering of layers within nanoflakes and bulk samples of $CuBiP_2Se_6$ and $CuInP_2S_6$, it is demonstrated that they also control the internal geometrical structure of isolated monolayers of these materials. This internal structure involves large displacements of the copper atoms, either normal to the layer plane or else within the plane, that change its ligation environment. In both cases, the van der Waals dispersion force out-competes traditional bonding effects to control structure. However, we find that the aspects of the dispersion force giving rise to each effect are uncorrelated: long range effects control inter-layer ferroelectric ordering whereas short-range effects control internal layer structure. These conclusions are drawn considering predicted properties of monolayers, bilayers, and bulk materials obtained using 14 density-functional-theory based methods. While the different methods used often predict starkly different quantitative results, they concur as to the basic nature of $ABP_2X_6$ materials. Of the methods used, only the PBE-D3 and optPBEvdW methods were found to predict a wide range of observed properties without serious disparity. Finding optimal computational methods remains a significant challenge for which the unusual multi-scale nature of the van der Waals interactions in $ABP_2X_6$ materials provides demanding criteria.

PACS Numbers

31.13.E dft  or  31.15.eg  functionals

68.65.ac  multilayers

77.80.-e   ferroelectric - antiferroelectric

I.   INTRODUCTION

The competition between traditional chemical forces involving covalent and ionic bonding with the van der Waals (vdW) force is being recognized as a widespread phenomenon [1]. Many different computational procedures are now available for modeling this competition, and we utilize these to consider the structure and properties of some 2D layered materials. These compounds are selected as the properties of the "van der Waals gap" is of critical current concern as a controlling influence on the properties of many conceived devices [2], but we here show that equally important in layered materials is also the effect the force has on the internal properties of each layer.

Some metal seleno- and thio-phosphate materials $ABP_2X_6$ (A, B = metals with valences either A(I)B(III) or A(II)B(II); X = S or Se) form lamellar structures in which thin molecular layers (see Fig. 1) self-assemble under the influence of the van der Waals force [3]. Each molecular layer involves $(X_3P-PX_3)^{4-}$ anions of near $D_{3d}$ symmetry, assembled with their P-P bonds normal to the layer. In this way, two parallel "sheets" of S or Se atoms, that are of order 3.5 Å apart, form in a hexagonal lattice, with the metal atoms locating in octahedral holes between the sheets. Figure 1 depicts such a high-local-symmetry structure, named "O". It manifests $S_6$ improper rotation axes normal to the layer that pass through the A atom, the B atom, and the P-P bond, as well as three $C_2$ axes bisecting them. The $C_2$ axes are important as, within a monolayer, they make the top sheet of S or Se atoms symmetrically equivalent to the bottom sheet.

If A and B are different atoms, then the O structure cannot have the full symmetry of the lattice. Indeed, it represents a high-energy unstable structure on the potential-energy surface, leading to 8 polymorphic forms associated with displacements of, in particular, the small Cu(I) atom. It can either move vertically ($z$ direction) towards one of the two layer sheets or horizontally in either direction along one of the three $C_2$ axes. Vertical displacements cause the loss of all $C_2$ operators and reduce $S_6$ to $C_3$. Horizontal displacements loose the $S_6$ operator altogether but preserve one of the $C_2$ operators, here taken to be the one in the $x$ direction.

Of considerable interest is the dipole polarization associated with internal displacements within each cell. The (hypothetical) O structure has inversion symmetry and hence manifests no dipolar polarization, but the magnitude of internal displacements can be quite large, of order 1.7 Å normal and 0.5 Å horizontal, so inducing significant electric polarization within a cell. These effects are sketched in Fig. 2a.

A monolayer is made up of a 2D grid of cells; its macroscopic properties will therefore be controlled by the nature of any short-range or long-range ordering of individual cell dipoles. Monolayers of cells displaying vertical displacements that align either ferroelectrically or else randomly have been observed [4]. However, monolayers of cells with horizontal displacements are yet to be observed. As a result, discussion of ferroelectricity has historically focused on polarization normal to the monolayer, and indeed we follow this lead. Hence in Figs. 1 and 2b we name the vertically displaced fully-ordered monolayer as *ferroelectric* (F). Shown also in these figures is an ordered *antiferroelectric* (A) structure based on vertical displacements, and variants with random (R) polarization are known. In contrast, horizontally displaced structures preserve one $C_2$ operator and hence there is no *z*-dipole component by symmetry, so we name these structures as *paraelectric*; two such paraelectric forms are possible, one named $P_s$ in which the A and B atoms move together to a short distance, ca. 3 Å apart, that can be considered as a (somewhat long) chemical bond, and one named $P_l$ in which they move apart to display a long unique interaction distance, see Figs. 1 and 2b. We show interest in such structures as one day they may be observed in some system, possibly with an exploitable metal-metal bond. More pertinently, however, their properties could shed considerable light on the interaction between ionic and van der Waals forces in controlling monolayer structure. To encompass such possibilities, our notation here is more general than that found elsewhere in which vertically displaced randomly aligned cells are sometimes referred to as being "paraelectric" [4].

Nanoflakes and bulk samples of $ABP_2X_6$ materials also display inter-layer ordering. For example, monolayers that are internally ferroelectrically ordered may be stacked together either ferroelectrically or antiferroelectrically. Materials that have been previously investigated for their net ferroelectric properties include: $CuInP_2S_6$ [3, 5-9] $CuBiP_2Se_6$ [4], $AgBiP_2Se_6$ [4], $AgBiP_2S_6$ [4], $CuCrP_2S_6$ [10], and $CuVP_2S_6$ [11]. $CuBiPS_2Se_6$ is observed at 97 K to form into an antiferroelectric polymorph comprised of vertically-displaced internally ferroelectrically-ordered monolayers stacked with alternating polarization, with dipole ordering being lost as the temperature rises to 173 K and then to 298 K [4]. It has been suggested that random loss of dipole ordering within a monolayer is responsible for this effect [4], but alternate possibilities include the formation of antiferroelectric monolayers as well as the formation of paraelectric monolayers (see Figs. 1 and 2). In contrast, $CuInP_2S_6$ was found to be ferroelectric with a Curie temperature of 315 K, thus sustaining a ferroelectric state up to room temperature [3, 5-7, 9, 12, 13]. Understanding the processes driving and opposing ferroelectricity in these materials therefore presents a significant modern challenge. In general, the origin of antiferroelectricity in materials remains poorly understood [14].

Recently, we demonstrated that, in $CuBiP_2Se_6$ and $CuInP_2S_6$, the van der Waals force controls ferroelectric ordering rather than traditional chemical forces [15]. This effect arises from strong dispersion interactions involving the copper ions of one internally ferroelectric monolayer interacting with the copper atoms in its neighbouring layers. Figure 2c depicts two fundamentally different types of antiferroelectrically arranged bilayers of such a monolayer. Most significantly, antiferroelectric nanoflakes and bulk materials by necessity must comprise alternating arrangements of each type. They involve different copper-copper inter-layer interactions: in one case, named $A_{ii}$, the vertically displaced Cu atoms both appear on the inside of the two layers, while in the other case, named $A_{oo}$, the two copper atoms appear on the outside. Alternatively, bilayers made from ferroelectrically aligned monolayers, named F, (see Fig. 2c), present a single type of inter-layer interaction in which the copper atoms have intermediary separation to those in $A_{oo}$ and $A_{ii}$. Shorter copper to copper distances result in stronger van der Waals attraction, meaning that the dispersion force scales non-linearly in order $A_{ii} > F > A_{oo}$. Comparison of the average interlayer interaction in antiferroelectric structures with that in the ferroelectric one then controls ferroelectric ordering in the crystals. These net contributions to the total energy remain much larger than the electrostatic energy differences associated with both short-range and long-range dipole ordering. Hence the inter-layer van der Waals force controls the ferroelectric ordering of monolayers within bilayers, nanoflakes, and bulk solids of these materials.

To understand this result more, we require independent means of assessing the nature of the van der Waals forces, plus ways of assessing the nature of the intrinsic chemical and electrostatic forces. This includes both qualitative methods for the interpretation of observed results as well as quantitative schemes for the *a priori* prediction of the effects of the forces involved.

The chemical bonding forces that singularly control ferroelectricity in traditional three-dimensional materials are mostly well described using density-functional theory (DFT), employing simplistic functionals based on the generalized-gradient approximation (GGA) such as the Perdew, Burke, and Ernzerhof (PBE) [16] functional. Indeed, this method has been used to predict the ferroelectric order in lamellar materials related to our systems of interest [4, 9, 17], including the prediction of spontaneous dipole ordering in $CuInP_2S_6$ [9]. Often semiconductors with trigonal, tetrahedral and octahedral symmetry experience off-center atomic displacements owing to the presence of partially occupied orbitals [18]. Related off-center structural instabilities have also been investigated in various perovskite $ABO_3$ ferroelectric compounds such as iron defect sites in $KTaO_3$ [19], $Ba_{0.5}Sr0.5TiO_3$ [20], and $BaTiO_3$ [21]. In such bulk ferroelectric compounds, ferroelectric instability results from a competition between long-range Coulomb interactions favoring the ferroelectric polymorph and short-range forces supporting the undistorted paraelectric structure. Properties are then determined by the lattice dynamics [22] in competition with the electron screening of the Coulomb interactions [21]. Most relevant, however, is the analysis of $CuBiP_2Se_6$ by Gave et al. [4] which explains the off-centre displacement of the Cu(I) ion in terms of a second-order Jahn-Teller distortion.

Naively, one would expect the PBE density functional to provide a reasonable description of all of these chemical effects. However, improved computational methods are available, including the extension of PBE to make the HSE06 hybrid density functional [23]. This allows for a much more realistic description of the electron exchange effects that often control chemical reactions and ferromagnetism.

In contrast, neither GGA nor hybrid functionals provide a realistic description of the dispersion force, the force responsible for the attractive part of the van der Waals interaction. The last decade has seen the development of many methods that take into account this basic deficiency in DFT methodology [24]. Indeed, in the area of materials research, density-functionals with van der Waals corrections have been applied to critical questions such as the prediction of interlayer binding in various two-dimensional heterostructures including graphene, hexagonal boron nitride, phosphorene, and transition metal dichalcogenides [25]. Our previous demonstration [15] that van der Waals forces control ferroelectric ordering in $ABP_2X_6$ materials builds on this research.

Looking beyond this widely accepted understanding of the significance of van der Waals forces, today, many examples of systems traditionally believed to be treated well using DFT without van der Waals correction are now being found to be systems in which this effect is actually critical [1]. Noteworthy is that the dispersion can be large for interactions involving soft atoms like S, Se, In, and Bi, with the implication that it could also control atomic structure *within* a single monolayer. $ABP_2X_6$ materials therefore present a unique situation in which the role of the van der Waals force in controlling intermolecular interactions can be studied in parallel with its role in competing with chemical forces to control local bonding. This in turn provides a unique opportunity to assess modern computational approaches for modelling the van der Waals force.

Just as many different density functionals have been developed, so also have many van der Waals approaches. To date, the self-assembly of $ABP_2X_6$ layers has been considered using only one such approach, correcting the PBE density functional using the "D2" empirical dispersion correction of Grimme et al. [26] (PBE-D2), to investigate ferroelectric ordering in $CuIn_2P_3S_9$ and $CuInS_2P_6$ [7]. For these systems, PBE-D2 was shown to give similar ordering to that obtained using PBE [9]. Here, we consider pure PBE as well as 11 different combinations of van der Waals corrections with PBE-based GGA density functionals. We also consider the HSE06 functional, and in addition, that combined with the "D3" empirical dispersion correction of Grimme et al. [27, 28].

While GGA and hybrid functions are naively thought to provide a good description of electron correlation at short distance and van der Waals corrections thought to provide a good description at long distance, no clear-cut separation really exists, with both contributions applicable to a broad intermediary region. For example, the crude local-density approximation (LDA) to DFT provides an exact description of the dispersion force at *all* length scales in a free-electron gas. Changing the density functional modifies how a van der Waals correction describes long-range interactions, while changing a long-range correction also modifies chemical bonding forces. Hence each density functional and applied correction present a combination that should be considered as a unique methodology. Fourteen computational methods were used, as summarized in Table I. It is the long-range aspect of each combination that dominates traditional areas of van der Waals force research, such as those demonstrated to control self-assembly in general and also specifically ferroelectric ordering in $ABP_2X_6$ materials [15], while the short to medium range part controls the competition between van der Waals and covalent bonding [1].

These 14 computational methods are applied to study first monolayers, then bilayers, then bulk solids of $CuBiP_2Se_6$ and $CuInP_2S_6$. Looking only at monolayers allows not only their basic properties (Fig. 1) to be investigated at a fundamental level, but also how these phenomena are perceived using the different computational methods that embody the van der Waals force. Progressing to bilayers then allows the traditional inter-layer aspect of the van der Waals force to be depicted. Moving to solids then allows assessment of the computational methods, and the visions they portray concerning the interplay of van der Waals and chemical forces, to be examined in the light of a range of observed phenomena.

## II. COMPUTATIONAL METHODS

All calculations were performed using VASP 5.4.1 [29], where the valence electrons were separated from the core by use of projector-augmented wave pseudopotentials (PAW) [30]. The energy cut-off for the plane-wave basis functions was set at 500 eV. The energy tolerance for the electronic structure determinations was set at $10^{-7}$ eV to ensure accuracy. We used **k**-space grids of 8×8×1 for the GGA functionals and 4×4×1 for HSE06, with test calculations indicating that the reduced sampling introduces relative errors in calculated energies of order 3 meV. VASP performs periodic imaging in all three dimensions and so to model monolayers and bilayers we introduced a vacuum region of ca. 15 Å between the periodic images in the direction normal to the layer(s). To minimize interactions between these periodic images, we applied the dipolar correction implemented in VASP (by setting the "IDIPOL" to "TRUE"). However, for a selection of structures, we tested the influence of this correction, finding that it does not influence the qualitative or quantitative nature of the results.

Geometry optimizations were made for all internal structures, terminating when the forces on all atoms fell below 0.01 eV/Å. In monolayer and bilayer calculations, the in-plane unit-cell dimension $a$ was kept fixed at the value observed for the related bulk materials of 6.6524 Å for $CuBiP_2Se_6$ [4] and $a$=6.0955 Å, $b$ = 10.5645 Å and $\beta$ = 107.101° for $CuInP_2S_6$ [12], with the $CuInP_2S_6$ calculations done enforcing hexagonal symmetry. The starting structures for geometry optimizations were based on the properties of adjacent layers observed in these materials, with test calculations done for alternate structures using the PBE-D3 method not realizing lower-energy alternatives. For bulk structures, all lattice vectors were fully optimized.

The 14 computational methods considered, (see Table I), include the raw PBE density functional, plus this combined with various dispersion corrections: Grimme's D2 empirical correction [26] (PBE-D2), Grimme's D3 empirical correction [27] in its original form without Becke-Johnson damping with PBE (PBE-D3) and revPBE [31] (revPBE-D3), the exchange-hole based correction of Steinmann and Corminboeuf [32] (PBE-dDsC), the Tkatchenko-Scheffler method [33] (PBE-TS), that with self-consistent screening (SCS) [34] (PBE-SCSTS), and that extended to make Tkatchenko's many-body dispersion method [35, 36] (MBD@rsSCS) (VASP flag "IVDW= 202"). The other computational methods were all based around the pioneering van der Waals density-functional approach devised initially by Lundqvist et al. [24, 37]. This correction was applied to the revPBE density functional [38] (revPBEvdW), the optPBE density functional [38] (optPBEvdW) or the optB88 density functional [38] (optB88-vdW). Also, in a form modified by Lee *et al.*, it was combined [39] with the BP86 density functional [40] (vdW-DF2). Finally, we also considered the raw HSE06 hybrid density functional [23] as well as that combined with D3 [28].

### III. RESULTS AND DISCUSSION

Optimized Cartesian coordinates for monolayers, bilayers, and bulk solids calculated using the different computational methods are reported in full in Supporting Data. Extensive summaries of system properties are also provided therein, with particularly important properties also reported in the main text figures and tables. All computational methods considered predict that $CuBiP_2Se_6$ and $CuInP_2S_6$ materials are narrow band-gap semiconductors, with for example the PBE bandgaps being of the order of 1 eV.

### A. van der Waal forces control the internal structure of isolated $ABP_2X_6$ monolayers

Figures 1 and 2b sketch the basic chemical processes available to $ABP_2X_6$ monolayers as their cells distort from their (hypothetical) octahedral structure O. We do not consider explicitly random structures R of any form, focusing instead on ferroelectric monolayers F, antiferroelectric monolayers A, and the two possible types of paraelectric monolayers $P_s$ and $P_l$. For the 14 computational methods used, Table II compares key values for the Cu to In or Bi displacements in the $z$ and $x$ directions, $\Delta z$ and $\Delta x$, respectively, the vertical displacement between the Cu atoms in the antiferroelectric monolayers $\Delta z_{CuCu}$, and the relative energy difference per cell of the A, $P_s$, and $P_l$ structures compared to F, $\Delta E^F$; additional related properties are listed in Supporting Data Table S1.

The vertical displacement data is highlighted in Fig. 3. First, $\Delta z$ is compared for ferroelectric monolayers of $CuBiP_2Se_6$ and $CuInP_2S_6$ in Fig. 3a. In this and later figures, results for each computational method are presented using the hexadecimal code "0" to "D", as defined in Table I. Different computational methods predict variations in $\Delta z$ of up to 0.8 Å, indicating that the internal structure of monolayers in isolation is highly sensitive to the treatment used for the van der Waals force and its associated GGA or hybrid density functional. $CuInP_2S_6$ is found to be only 73% as sensitive to these effects as is $CuBiP_2Se_6$, a reasonable result given that the van der Waals force is naively expected to be stronger in $CuBiP_2Se_6$ as Bi is softer than In and Se is softer than S. Figures 3b and 3c compare $\Delta z$ in the ferroelectric monolayer to $\Delta z_{CuCu}$ in the antiferroelectric one, demonstrating a linear correlation for $CuInP_2S_6$ but a bifurcated pattern for $CuBiP_2Se_6$, suggesting that the van der Waals force is more perturbative for $CuInP_2S_6$ and more dominant for $CuBiP_2Se_6$.

Returning directly to Table II, we consider the displacements $\Delta x$ occurring in the paraelectric monolayers $P_s$ and $P_l$. For $CuInP_2S_6$, the structures $P_s$ present a short apparent bond between the Cu and In atoms of length ~ 2.9 Å. This length is contracted from the value of 3.52 Å found in the ferroelectric monolayer by 0.53 to 0.72 Å, depending on the computational method. For $CuBiP_2Se_6$, the situation is similar: the apparent Cu to Bi bond length is ~ 3.1 Å, contracted from the value of 3.78 Å found in the ferroelectric monolayer by 0.62 to 0.76 Å. Greater variation is found in the alternate paraelectric structures $P_l$ that manifest one unique long inter-metallic interaction, the increase in interaction distance being 0.07 to 0.28 Å for $CuInP_2S_6$ and 0.13 to 0.55 Å for $CuBiP_2Se_6$. From the energy differences $\Delta E^F$ reported in the table, we see that most computational methods predict $P_s$ to be more stable than $P_l$. For $CuBiP_2Se_6$, all methods predict this result with a preference of 27 to 73 meV, but for $CuInP_2S_6$, PBE-D2 and HSE06-D3 predict $P_l$ to be the most stable, with a much wider range in predicted energy differences. In summary, we see that different combinations of van der Waals corrections with density functionals have smaller effects on geometrical structures for displacements in the $x$ direction than they do for displacements in $z$, but nevertheless key energetic properties remain determined by these factors.

Focusing more on the energy differences, the paraelectric structures are all predicted to be much less stable than the ferroelectric and/or antiferroelectric ones, consistent with the lack of observation of paraelectric species. This energy preference is by 170 to 315 meV for $P_s$ of $CuInP_2S_6$, but by values as low as 60 meV for $CuBiP_2Se_6$. Increasing the van der Waals interactions therefore favours paraelectric forms with short A-B distances and so it is conceivable that some $ABP_2X_6$ species could one day be made. Exploitation of A-B spin or other interactions may make such species of practical interest.

For $CuInP_2S_6$, all methods predict that the antiferroelectric monolayer is more stable than the ferroelectric one by 6 to 46 meV. This is inconsistent with the observation of only ferroelectric structures for this compound. However, monolayers themselves are yet to be observed. The observed ferroelectric bulk and nanoflake structures [3, 5-9] may result as a consequence of inter-layer interactions, given that many methods predict only small energy differences less than 20 meV. For $CuBiP_2Se_6$ with its stronger van der Waals interactions, the computational methods predict energy differences between the antiferroelectric internal monolayer structure and the ferroelectric one ranging from -70 to 31 meV. Both types of structures are observed in $CuBiP_2Se_6$ crystals [4], indicating small energy differences. These issues are pursued in subsequent subsections.

## B. van der Waals forces and the properties of $ABP_2X_6$ bilayers

Table III lists the calculated energies of the antiferroelectric bilayers $A_{ii}$ and $A_{oo}$ made from ferroelectric monolayers, and well as energies for bilayers MA made from antiferroelectric monolayers, relative to the energies of the ferroelectric bilayers F. In addition, the calculated vertical displacements between the Cu and In or Bi atoms, $\Delta z$, are listed in Supporting Data Table S1, with also the inter-layer spacings between the monolayers of the bilayer, $d$, listed in Supporting Data Table S2. Further, the data from Tables II and III is represented in a different way in Supporting Data Table S3 as the energy of binding $\Delta E$ of the two monolayers to make the bilayer. This information, along with results from Table II, is correlated in Figs. 5 and 6: Fig. 5 compares average energies of binding $\Delta E$ for $CuInP_2S_6$ versus those for $CuBiP_2Se_6$, whereas Fig. 6 compares average $\Delta E$, $\Delta z$, and $d$ values for each individual compound. These averages are evaluated considering properties of the $A_{ii}$, $A_{oo}$, MA, and F bilayers, with expanded results depicting each individual bilayer type shown in Supporting Data Figs. S1-S5. While the expanded data reveal significant systematic differences between results obtained for the various bilayer types, the features of greatest relevance herein are independent of these perceived variations, being generic properties of inter-layer interactions rather than specific structure-related effects.

First, we consider the average energies of binding, $\Delta E$, of the monolayers. These are indicative of the strength of the classic van der Waals attractive force that holds together lamellar materials, biological structures, polymers, etc.. Figures 5 and S1, also Table S3, show that most methods predict similar results for $CuInP_2S_6$ as for $CuBiP_2Se_6$. The binding for $CuBiP_2Se_6$ is typically predicted to be 50 to 100 meV stronger than that for $CuInP_2S_6$, consistent with the general expectation that the van der Waals forces are stronger within $CuBiP_2Se_6$. There are some exceptions, however, with results ranging from -50 to 250 meV; PBE-D2 predicts the greatest enhancement in binding while PBE-SCSTS depicts the reversed effect. In absolute magnitude, the binding energies range from -400 to -800 meV per cell, all indicative of strong binding yet the total magnitude shows considerable variation. PBE-SCSTS binds the strongest, whereas PBE-TS binds the weakest.

Next, Fig. 6a-b shows that the vertical displacements $\Delta z$ between the Cu and In or Bi atoms calculated for the bilayers strongly correlate with the values predicted for the ferroelectric monolayer. This indicates that internal layer polarization is controlled by the van der Waals forces acting within each layer and is not sensitive to van der Waals forces acting between layers. Further, Figs. 6c-h show that there is no correlation between the bilayer interaction energy $\Delta E$ or the bilayer inter-layer spacing $d$ with $\Delta z$, indicating again that the computational methods perceive the van der Waals forces acting between layers very differently to how they perceive them acting within layers. These layered $ABP_2X_6$ materials hence present a scenario in which the critical component is electron correlation at length scales intermediary between the short-range part, which is normally well treated with GGA-type density functionals, and the long-range part, which is normally well treated by DFT dispersion correction approaches.

Of interest also is the magnitude of the optimized inter-layer spacings $d$. The different van-der-Waals corrected methods used predict values over a substantial range, from 3.1 to 3.5 Å. As expected, the computational methods used that do not include van der Waals corrections, PBE and HSE06, predict much larger values than those that do include them (Table S2). PBE does not predict stable bilayers for the antiferroelectric bilayers of $CuInP_2S_6$ (Table S3) and hence the reported spacings $d$ merely reflect the convergence criterion set for the force minimization during the geometry optimization. Thus these methods fail severely for the evaluation of the interlayer interactions, as expected. Of note, however, are the values predicted by these methods for the short-range property $\Delta z$, which fall mid-way within the range predicted by the van-der-Waals corrected methods. The van der Waals corrections, with their associated modifications to the GGA density functionals, therefore seemingly perturb the short-range correlation in a random way, sometimes enhancing it and sometimes reducing it.

### C. Assessment of the computational methods through examination of predicted bulk properties

Table IV lists calculated enthalpies $\Delta H$ for various polymorphs of bulk $CuInP_2S_6$ and $CuBiP_2Se_6$, while Table V compares related calculated geometric properties with observed values; results are provided for only the 12 computational methods not involving hybrid functionals. The structures considered are shown in Fig. 7. These are all periodic after 6 layers in the direction normal to the plane, but some have only two unique layers per cell. They include the observed low-temperature structure for each compound, as well as that modified by pushing the Cu atoms in every second layer to the other side of their octahedral cavity (interchanging ferroelectric and antiferroelectric arrangements) [15]. Other variants are also considered, inspired by scanning the bilayer configuration space using the PBE-D3 method in search of alternative low-energy arrangements. Two primary issues are considered: the ability of the different computational methods to predict the actual observed polymorphs, and their ability to predict accurately the structural details of those polymorph.

In Table IV, the calculated enthalpies at atmospheric pressure are listed with respect to those for the observed low-temperature polymorphs; the associated relative interaction energies $\Delta E$ are listed in Supporting Data Table S4 and are very similar. If the reported enthalpy difference is negative, then the calculation predicts a result that is contrary to low-temperature experimental observation. Only two methods correctly predict the observed structure of both materials: revPBEvdW and vdwDF2. The data is summarized in the column in the table that lists the total number of negative $\Delta H$ values predicted by each method, i.e., the number of structures considered that are predicted to be more stable than the observed structure. However, many enthalpy differences are quite small, indicating in general that many polymorphs are likely to have similar enthalpies, making the observed structure intrinsically difficult to predict. To allow for this, the last column in the table lists the total number of predictions with a significant enthalpy error of magnitude > 10 meV. Only three methods fail this less-demanding criterion: revPBE-D3, PBE-SCSTS, and PBE-MBD@rsSCS.

Similarly, Table V shows deviations from experiment for the observed polymorph pertaining to: the *a* axis vector length, the periodicity dimension normal to the layers, $c_z$, the interlayer distances across antiferroelectric boundaries ($d_{ii}$ or $d_{oo}$) and ferroelectric boundaries ($d_{io}$), as well as the vertical separation between Cu and In or Bi atoms, $\Delta z$. Summarizing the results, for each method, the number of properties for which there are only small deviations and the number for which there are large deviations are counted in the right-hand columns of the table. This reveals that PBE-D2 and optB88vdW perform the best for quantitative geometrical predictions. However, a method that performs poorly in Table IV for enthalpies, revPBE-D3, performs very well in this regard, whereas conversely the methods that perform best for enthalpies, revPBEvdW and vdwDF2, perform the worst of all van der Waals corrected approaches for geometrical properties. This result is consistent with our recent comparison of van der Waals functionals in broader contexts [41].

## D. Overall assessment of the computational methods

Tables IV and V depict how the 12 computational methods not involving hybrid functionals fare in terms of predicting observed properties of $CuInP_2S_6$ and $CuBiP_2Se_6$ crystals. However, not included therein are related properties of crystals containing monolayers with internal random or antiferroelectric order, owing to feasibility issues associated with the very large size of pertinent samples. Nevertheless, Tables II and III indicate that such neglected considerations are important in assessing the performance of different methods. Hence in Table VI we combine the summary information from Tables III-IV, taking computed bilayer properties as being indicative of nanoflakes and bulk solids that embody them. Table VI indicates, for each computational method, whether or not a serious failure was identified in Tables IV-V or else implied in Table III. No such serious failures are found for only PBE-D3 and optPBEvdW. Of particular note is that PBE-MBDrsSCS, a method widely noted [24] as performing well for van der Waals interactions over a very wide range of properties, including situations in which the van der Waals forces cannot be described in terms of (fundamentally) pairwise-additive contributions, is the only approach considered that shows significant failures in all categories considered herein.

## IV. CONCLUSIONS

The results presented allow two different types of conclusions to be drawn: conclusions pertaining to the fundamental nature of $ABP_2X_6$ materials, and those pertaining to assessment of the most appropriate computational method for making quantitative property predictions. In the first case, conclusions flow from considering both (i), the properties predicted robustly by all van der Waals computational approaches, and (ii), the substantial differences that can occur in predicted properties and how these differences correlate one to another. We consider these three aspects separately.

Recognized earlier [15] is that the van der Waals forces control interlayer ferroelectric ordering in nanoflakes and solids of $CuInP_2S_6$ and $CuBiP_2Se_6$. This occurs as the interlayer van der Waals forces, which are significantly influenced by the embodied Cu-Cu interactions, scale non-linearly in magnitude for bilayer junction types in order $A_{ii} > F > A_{oo}$. Hence as antiferroelectric crystals involve junctions of both $A_{oo}$ and $A_{ii}$ types, partial cancellation of the van der Waals contributions gives a net bonding effect that can favour either the ferroelectric or the antiferroelectric forms, with the magnitude of this effect being typically much greater than the electrostatic interactions that favour ferroelectric structures. The calculated enthalpies for all the variant structures and computational methods listed in Table IV embody this qualitative effect, the details of which are reflected in the bilayer energy differences listed in Table III. Hence we see that, while different computational methods present significantly different results in terms of relative polymorph energies and geometrical properties, the basic qualitative scenario describing how ferroelectricity in $ABP_2X_6$ monolayers is controlled remains robustly described.

Substantial differences predicted for the internal structures of isolated $ABP_2X_6$ monolayers indicate that the van der Waal force, and in particular how it merges into traditional short-range descriptions of electron correlation forces, as manifested using various density functionals, controls internal monolayer structure. These contributions overpower those from the traditional forces associated with covalent and/or ionic bonding. This provides another example of the modern realization that van der Waals forces can out-compete traditional ones to control chemical bonding [42, 43].

Hence two major effects of van der Waals forces in controlling the structure of $ABP_2X_6$ monolayers, nanoflakes, and materials are identified: one acting within layers, the other acting between layers. The calculations show no correlation between how different van der Waals computational methods perceive these two types of effects. Traditionally, van der Waals corrections have been optimized for the description of long-range intermolecular forces, but here we see that how they merge with traditional covalent forces at short range can be equally as important.

Only two computational approaches, PBE-D3 and optPBEvdW, showed no major problems identified through quantitative comparison of calculated and observed properties. However, the research strategy used in this work is significantly biased against PBE-D3, as this method was used to seek out possibly competitive polymorphs; the other methods are therefore only assessed in regard to issues for which PBE-D3 is already identified as potentially failing. Repeating the calculations, searching for specific possible weaknesses in the other computational approaches, could therefore reveal additional undetected serious problems.

Mostly, the hybrid density functional methods HSE06 and HSE06-D3 performed similarly to their GGA counterparts PBE and PBE-D3. However, the differences were sometimes of the same order as those associated with variations in the van der Waals corrections. Hence it is not clear that GGA methods are in general sufficient for the robust treatment of $ABP_2X_6$ systems.

Considering all of the methods associated with Grimme-type empirical dispersion corrections (PBE-D2, PBE-D3, revPBE-D3, and HSE06-D3), for the examined properties of each method, at most one serious failure per method is reported in Table VI. This is in contrast to the results found for the Lundqvist family of methods: revPBEvdW, optB88vdW, optPBEvdW, and vdWDF2. As with PBD-D3, Table VI reports no serious failure for optPBEvdW, but revPBEvdW shows two out of three instances. In greater detail, the results presented for the Grimme methods have smaller ranges in Figs. 3 and 6 than those from the Lundqvist set. Results for the Tkatchenko-Scheffler series of methods PBE-TS, PBE-SCSTS, and PBE-MBD@rSCS show mostly the least variations in these figures (except for inter-layer binding energies), but they also show the most significant failures as summarized in Table VI.

The Tkatchenko-Scheffler series present the most successful efforts so far to include all commonly recognized important aspects of the dispersion interaction [44]. In terms of the fundamental physics included, variations within this set are greater than those within the Lundqvist and Grimme sets. Despite the shortcomings revealed herein, these approaches depict a significant direction forward for research into accurate modelling of the van der Waals force. In contrast, the Lundqvist family embody considerable effort in properly representing many of the long-range aspects of dispersion but fail to deal with the issue concerning how this long range part relates to the intrinsic GGA, relying on empirical combinations with available GGA variants to deliver useful results. Grimme's methods are the most empirical but rely on the use of typically realistic expressions for the long range part coupled with generally applicable merging strategies with the GGA part.

During the study of $ABP_2X_6$ materials, both the long-range and short-range aspects of van der Waals corrected density functions become important. Our results thus imply that of the Lundqvist-type methods, optPBEvdW appears the most reasonable empirical combination of the dispersion density functional with a GGA functional. Such a conclusion is hazardous, however, as in general this family delivers widely varying and unpredictable results when broadly applied [24]. From amongst the Grimme family and many other empirical schemes, revPBE-D3 has been identified as one of the best general approaches [45], yet for $ABP_2X_6$, PBE-D3 appears to be more reliable. Also, the most advanced Tkatchenko-Scheffler-type method considered, PBE-MBD@rSCS, predicts the poorest results. In any required application, careful choice of van der Waals correction remains a critical ongoing issue [24, 46].

# ACKNOWLEDGMENTS

This work was supported by resources provided by the National Computational Infrastructure (NCI), and Pawsey Supercomputing Centre with funding from the Australian Government and the Government of Western Australia, as well as Chinese NSF Grant #1167040630. This work is also supported by the Australian Research Council, ARC DP160101301.

**TABLE I:** The numbered computational methods used in this study, indicating the density functional used and its vdW correction.

| # | Method | functional | vdW |
|---|---|---|---|
| 0 | PBE | PBE | - |
| 1 | HSE06 | HSE06 | - |
| 2 | revPBE-D3 | revPBE | D3 |
| 3 | PBE-D2 | PBE | D2 |
| 4 | PBE-D3 | PBE | D3 |
| 5 | HSE06-D3 | HSE06 | D3 |
| 6 | PBE-dDsC | PBE | dDsC |
| 7 | PBE-TS | PBE | TS |
| 8 | PBE-SCSTS | PBE | PBE-SCSTS |
| 9 | PBE-MBD@rsSCS | PBE | MBD@rsSCS |
| A | revPBEvdW | revPBE | Dion |
| B | optB88vdW | optB88 | Dion |
| C | optPBEvdW | optPBE | Dion |
| D | vdWDF2 | PW86 | Dion-Lee |

**TABLE II:** Calculated properties of $CuBiP_2Se_6$ and $CuInP_2S_6$ monolayers, including the $z$ coordinate change between Cu and In/Bi atoms ($\Delta z$, in Å), the $z$ coordinate change between the two Cu atoms in the antiferroelectric structure ($\Delta z_{CuCu}$, in Å), the in-plane displacement between Cu and In/Bu ($\Delta x$, in Å), and the relative energies of other polymorphs with respect to the F polymorph ($\Delta E^F$, in meV per cell).

| Method | \multicolumn{6}{c}{$CuBiP_2Se_6$} | \multicolumn{6}{c}{$CuInP_2S_6$} |
|---|---|---|---|---|---|---|---|---|---|---|---|---|
| | $\Delta z^a$ | $\Delta z_{CuCu}$ | $\Delta x^b$ | | $\Delta E^F$ | | $\Delta z^a$ | $\Delta z_{CuCu}$ | $\Delta x^b$ | | $\Delta E^F$ | |
| | F | A | $P_s$  $P_l$ | | $P_s$  $P_l$  A | | F | A | $P_s$  $P_l$ | | $P_s$  $P_l$  A | |
| PBE | 1.71 | 1.84 | 3.03 | 4.18 | 60 | 121 -42 | 1.70 | 3.09 | 2.82 | 3.73 | 250 | 291 -32 |
| HSE06 | 1.60 | 2.57 | 3.16 | 4.19 | 56 | 129  23 | 1.65 | 2.96 | 2.91 | 3.71 | 232 | 233 -34 |
| revPBE-D3 | 1.27 | 1.44 | 3.14 | 4.21 | 24 | 56 -49 | 1.41 | 2.55 | 2.93 | 3.70 | 188 | 212  -6 |
| PBE-D2 | 1.40 | 1.58 | 3.07 | 4.19 | 54 | 80 -70 | 1.44 | 2.59 | 2.98 | 3.62 | 225 | 209 -25 |
| PBE-D3 | 1.57 | 2.67 | 3.04 | 4.09 | 84 | 148  27 | 1.64 | 2.93 | 2.83 | 3.64 | 269 | 311 -17 |
| HSE06-D3 | 1.62 | 2.64 | 3.16 | 4.07 | 78 | 111  30 | 1.65 | 2.96 | 2.85 | 3.79 | 257 | 253 -33 |
| PBE-dDsC | 1.59 | 1.72 | 3.03 | 4.11 | 19 | 74 -53 | 1.63 | 2.92 | 2.84 | 3.69 | 214 | 245 -19 |
| PBE-TS | 1.59 | 1.68 | 3.04 | 4.14 | 10 | 65 -59 | 1.63 | 2.91 | 2.84 | 3.70 | 183 | 212 -19 |
| PBE-SCSTS | 1.72 | 2.98 | 3.02 | 4.21 | 71 | 136  15 | 1.75 | 3.15 | 2.80 | 3.73 | 237 | 310 -46 |
| PBE-MBD@rsSCS | 1.72 | 2.96 | 3.05 | 3.91 | 113 | 166  31 | 1.75 | 3.13 | 2.84 | 3.80 | 315 | 351 -20 |
| revPBEvdW | 1.83 | 2.97 | 3.08 | 4.17 | 113 | 162  15 | 1.78 | 3.22 | 2.91 | 3.66 | 264 | 283 -22 |
| optB88vdW | 1.58 | 1.69 | 3.02 | 4.33 | -4 | 51 -62 | 1.63 | 2.90 | 2.84 | 3.62 | 171 | 194 -15 |
| optPBEvdW | 1.71 | 1.81 | 3.05 | 4.17 | 60 | 112 -35 | 1.69 | 3.10 | 2.88 | 3.66 | 225 | 246 -17 |
| vdWDF2 | 1.97 | 3.39 | 3.11 | 4.15 | 219 | 270  22 | 1.89 | 3.43 | 2.99 | 3.59 | 371 | 377 -10 |

a: zero as a result of $C_2$ symmetry in paraelectric structures
b: values resulting from the hexagonal symmetry of F structures: 3.783 Å for $CuBiP_2Se_6$, 3.515 Å for $CuInP_2S_6$.

TABLE III. Energy of alternate structures with respect to the ferroelectric one, in meV, for bilayers of $CuBiP_2Se_6$ and $CuInP_2S_6$.

| method | CuBiP$_2$Se$_6$ | | | | CuInP$_2$S$_6$ | | | Nber.[a] < -30 |
|---|---|---|---|---|---|---|---|---|
| | $\Delta E_{Aii}$ | $\Delta E_{Aoo}$ | $\Delta E_{MA}$ | $\Delta E_{MA} - (\Delta E_{Aii}+\Delta E_{Aoo})/2$ | $\Delta E_{Aii}$ | $\Delta E_{Aoo}$ | $\Delta E_{MA}$ | |
| PBE | -1 | 2 | -80 | -81 | 5 | 7 | -78 | 2 |
| HSE06 | -5 | 6 | 48 | 48 | -11 | 0 | -73 | 1 |
| revPBE-D3 | -63 | 56 | -18 | -15 | -81 | 73 | 8 | 0 |
| PBE-D2 | -42 | 35 | -69 | -66 | -42 | 41 | -38 | 2 |
| PBE-D3 | -86 | 61 | 30 | 43 | -92 | 88 | -22 | 0 |
| HSE06-D3 | -64 | 32 | 69 | 85 | -59 | 65 | -57 | 1 |
| PBE-dDsC | -36 | 6 | -74 | -59 | -33 | 44 | -36 | 2 |
| PBE-TS | -45 | 29 | -75 | -67 | -43 | 46 | -22 | 0 |
| PBE-SCSTS | -76 | 43 | 52 | 69 | -100 | 101 | -79 | 1 |
| PBE-MBD@rsSCS | -87 | 51 | 77 | 95 | -83 | 73 | -26 | 1 |
| revPBEvdW | -26 | 5 | 54 | 65 | -22 | 29 | -33 | 1 |
| optB88vdW | -40 | 17 | -65 | -54 | -40 | 42 | -13 | 1 |
| optPBEvdW | -31 | 12 | -22 | -13 | -30 | 38 | -22 | 0 |
| vdWDF2 | -23 | -7 | 64 | 79 | -26 | 36 | -4 | 0 |

[a]: number of entries in the CuBiP$_2$Se$_6$ $\Delta E_{MA}$ - ($\Delta E_{Aii}+\Delta E_{Aoo}$)/2 and CuInP$_2$S$_6$ $\Delta E_{MA}$ columns < - 30 meV, suggesting that antiferroelectric monolayers should dominate structures instead of the observed structures.

TABLE IV. Calculated polymorph enthalpies $H$, in meV per cell, relative to those for the observed crystal structures of CuBiP$_2$Se$_6$ and CuInP$_2$S$_6$ for various considered alternative layer alignments and ferroelectric orderings.

| Sample: | CuBiP$_2$Se$_6$ | | | | | | CuInP$_2$S$_6$ | | | | | | Nber. results[a] | |
|---|---|---|---|---|---|---|---|---|---|---|---|---|---|---|
| Alignment: | Obs.[b] | | Alt. 1 | | Alt. 2 | | Obs. | | Alt. 1 | | Alt. 2 | | < 0 | < -10 |
| Structure: | A | F | A | F | | | A | F | A | F | A | F | | |
| Symmetry: | $C_3, i$ | $C_3$ | $C_3, i$ | $C_3$ | | | $i$ | - | $i$ | - | $C_3, i$ | $C_3$ | | |
| PBE | [0] | -3 | 0 | 0 | | | 10 | [0] | 13 | 2 | 12 | 4 | 1 | 0 |
| revPBE-D3 | [0] | 5 | 25 | 0 | | | 5 | [0] | -15 | -11 | -14 | -23 | 4 | 4 |
| PBE-D2 | [0] | 2 | 35 | -7 | | | 11 | [0] | 5 | -3 | 6 | -9 | 3 | 0 |
| PBE-D3 | [0] | 7 | 28 | 2 | | | 11 | [0] | 5 | -2 | 5 | -6 | 2 | 0 |
| PBE-dDsC | [0] | 8 | 27 | 4 | | | 18 | [0] | 19 | -1 | 12 | -1 | 2 | 0 |
| PBE-TS | [0] | 5 | 31 | 11 | | | 13 | [0] | 11 | -1 | 12 | -2 | 2 | 0 |
| PBE-SCSTS | [0] | 5 | 16 | 9 | | | 14 | [0] | 1 | -8 | -3 | -17 | 3 | 1 |
| PBE-MBD@rsSCS | [0] | 6 | 15 | 10 | | | 9 | [0] | -8 | -12 | -13 | -24 | 4 | 3 |
| revPBEvdW | [0] | 4 | 7 | 8 | | | 14 | [0] | 15 | 0 | 12 | 0 | 0 | 0 |
| optB88vdW | [0] | 5 | 37 | -5 | | | 17 | [0] | 12 | -3 | 10 | -6 | 2 | 0 |
| optPBEvdW | [0] | 6 | 20 | 5 | | | 16 | [0] | 15 | -1 | 11 | -2 | 2 | 0 |
| vdWDF2 | [0] | 9 | 17 | 12 | | | 15 | [0] | 23 | 3 | 16 | 6 | 0 | 0 |

a: To adequately describe the experimental data, all reported numbers should be positive; the number of negative results, and also significant results < -10 meV are noted.

b: these structures are periodic after 6 layers in the $c$ direction, otherwise they are periodic after 2 layers.

TABLE V. Differences in calculated properties from the (listed) observed crystal data for $CuBiP_2Se_6$ and $CuInP_2S_6$, in Å.

| | $CuBiP_2Se_6$ | | | | | $CuInP_2S_6$ | | | | | Nber. results[a] | |
|---|---|---|---|---|---|---|---|---|---|---|---|---|
| | $a$ | $c_z$ | $d_{ii}$ | $d_{oo}$ | $\Delta z$[b] | $a$ | $b$ | $c_z$ | $d_{io}$ | $\Delta z$ | OK | large err. |
| obs | 6.55 | 13.25 | 3.07 | 3.13 | 1.48 | 6.09 | 6.09 | 12.97 | 3.17 | 1.75 | | |
| PBE | 0.09 | 1.70 | 0.82 | 0.70 | 0.21 | 0.08 | 0.08 | 1.59 | 0.70 | -0.05 | 4 | 5 |
| revPBE-D3 | 0.06 | 0.12 | -0.01 | 0.06 | -0.11 | 0.04 | 0.04 | 0.17 | 0.03 | -0.22 | 8 | 0 |
| PBE-D2 | 0.01 | 0.03 | -0.04 | 0.01 | -0.03 | 0.01 | 0.01 | 0.20 | 0.04 | -0.22 | 9 | 0 |
| PBE-D3 | 0.04 | 0.18 | 0.02 | 0.05 | 0.16 | 0.02 | 0.02 | 0.24 | 0.05 | -0.01 | 8 | 0 |
| PBE-DFTdDsC | 0.01 | 0.25 | 0.08 | 0.10 | 0.12 | 0.01 | 0.01 | 0.25 | 0.06 | -0.01 | 7 | 0 |
| PBE-TS | 0.07 | 0.49 | 0.20 | 0.12 | 0.30 | 0.05 | 0.05 | 0.43 | 0.14 | 0.10 | 4 | 2 |
| PBE-SCSTS | 0.08 | 0.57 | 0.24 | 0.15 | 0.29 | 0.06 | 0.06 | 0.26 | 0.05 | 0.08 | 5 | 0 |
| PBE-MBDrsSCS | 0.03 | 0.34 | 0.10 | 0.13 | 0.17 | 0.01 | 0.01 | 0.32 | 0.09 | 0.01 | 6 | 1 |
| revPBEvdW | 0.23 | 1.15 | 0.44 | 0.36 | 0.29 | 0.18 | 0.18 | 0.89 | 0.30 | 0.06 | 1 | 4 |
| optB88vdW | 0.04 | 0.10 | -0.01 | 0.00 | 0.12 | 0.03 | 0.03 | 0.11 | -0.02 | 0.03 | 9 | 0 |
| optPBEvdW | 0.13 | 0.56 | 0.19 | 0.14 | 0.22 | 0.10 | 0.10 | 0.48 | 0.13 | 0.04 | 3 | 0 |
| vdWDF2 | 0.29 | 0.98 | 0.32 | 0.21 | 0.38 | 0.22 | 0.22 | 0.74 | 0.19 | 0.17 | 0 | 4 |

a: OK results are taken to be those within ± 0.2 Å for $c_z$ and ± 0.1 Å otherwise, whereas those with large errors are taken to be those without 0.6 Å for $c_z$ and ± 0.3 Å, respectively.
b: The observed crystal structures are for mixtures of polymorphs and so the observed value is likely to be underestimated; this data is not included in the results summary.

TABLE VI. Summary of significant errors identified in calculation predictions (from Tables III-V).

| obs | MA suggested | Crystal geometries | Crystal energetics |
|---|---|---|---|
| PBE | x | | x |
| HSE06 | x | - | - |
| revPBE-D3 | | x | |
| PBE-D2 | x | | |
| PBE-D3 | | | |
| HSE06-D3 | x | - | - |
| PBE-dDsC | x | | |
| PBE-TS | | | x |
| PBE-SCSTS | x | x | |
| PBE-MBD@rsSCS | x | x | x |
| revPBEvdW | x | | x |
| optB88vdW | x | | |
| optPBEvdW | | | |
| vdWDF2 | | | x |

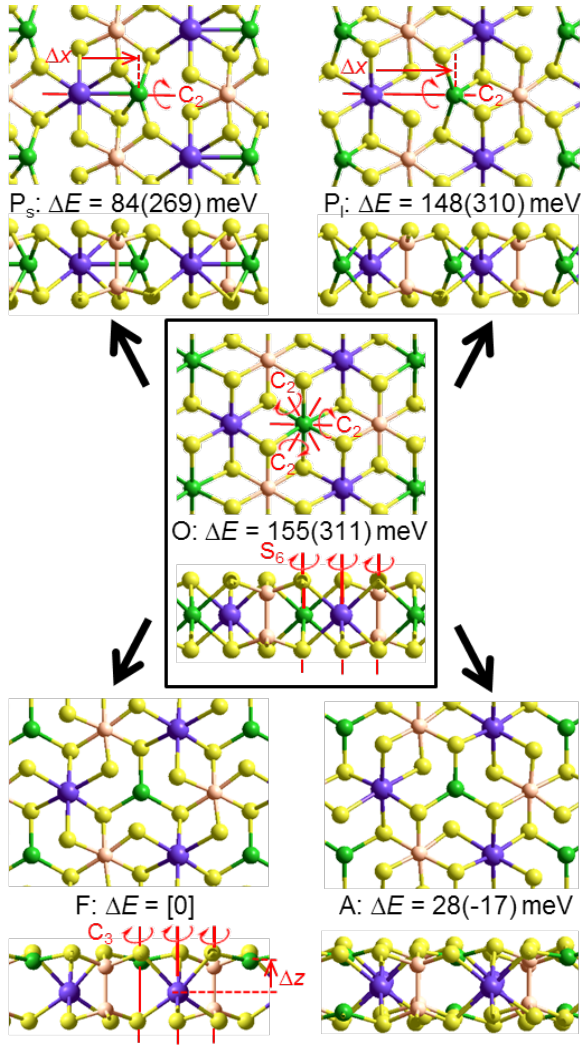

**FIG. 1.** (Color online) Top and sectioned side views of $ABP_2X_6$ monolayers, where A = Cu is in green, B = Bi or In is in magenta, X = S or Se is in yellow, and P is in orange, in a hypothetical high symmetry structure (O) and that after in-plane distortions of the Cu atom to make paraelectric ($P_s$, $P_l$) structures or else out-of-plane distortions to make polarized structures with internal long-range ferroelectric (F) or antiferroelectric (A) order. Highlighted are in-plane axes of local $C_2$ symmetry and normal axes of either $S_6$ or $C_3$ symmetry about the Cu atoms, as well as the normal ($\Delta z$) and in-plane ($\Delta x$) displacements of the Cu atom with respect to the Bi or In atom, in Å. The shown structures and energies per cell were optimized using PBE-D3 for $CuBiP_2Se_6$, with, in (), analogous energies for $CuInP_2S_6$.

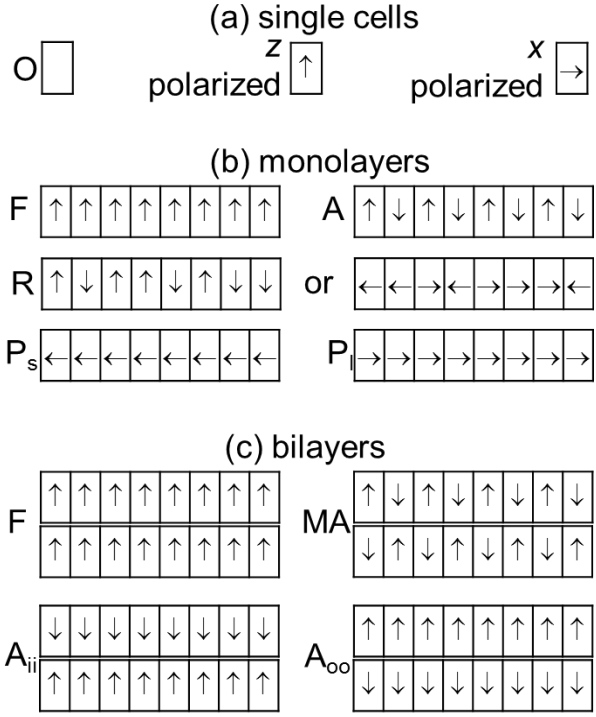

**FIG. 2.** Schematic representation of the possible ferroelectric (F), antiferroelectric (A), random (R), and paraelectric (P) dipole polarizations in single cells, monolayers, and bilayers of $ABP_2X_6$ compounds; structure MA is a bilayer with inversion symmetry made from antiferroelectric monolayers. The arrows indicate the displacement direction of the Cu atoms from the octahedral site (O), see Fig. 1 for atomic-scale details.

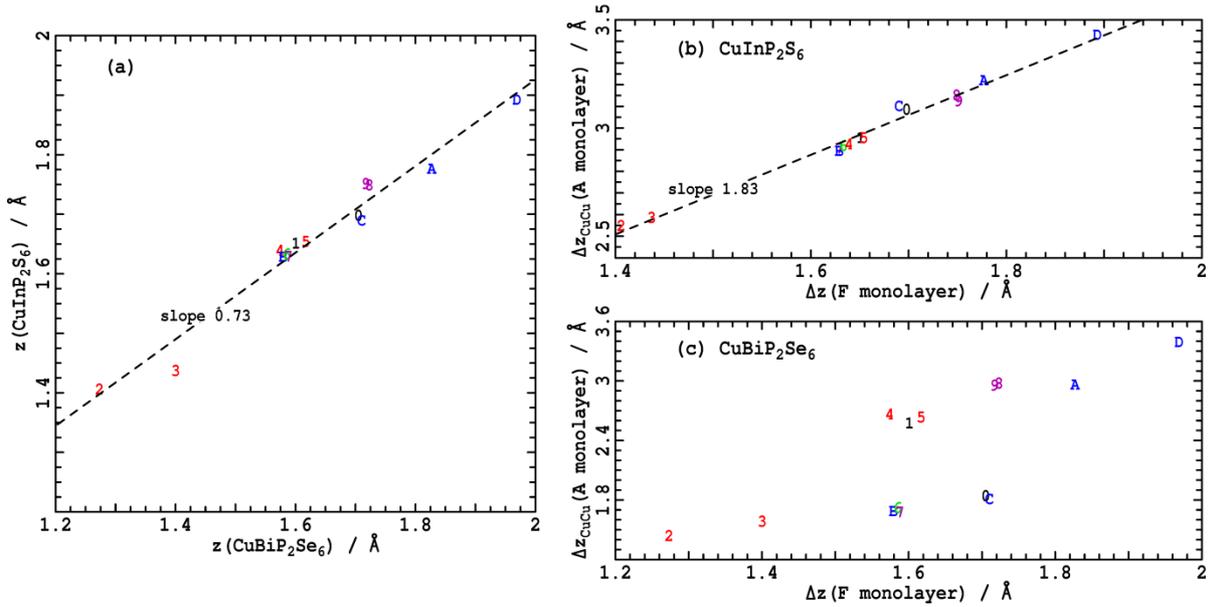

**FIG. 3.** Comparison of the calculated Cu to Bi or In distances $\Delta z$ for ferroelectric monolayers for $CuInP_2S_6$ and $CuBiP_2Se_6$, (a), and these quantities compared to the Cu-Cu vertical distance $\Delta z_{CuCu}$ for antiferroelectric monolayers, (b)-(c), for the various computational methods 0-D defined in Table I.

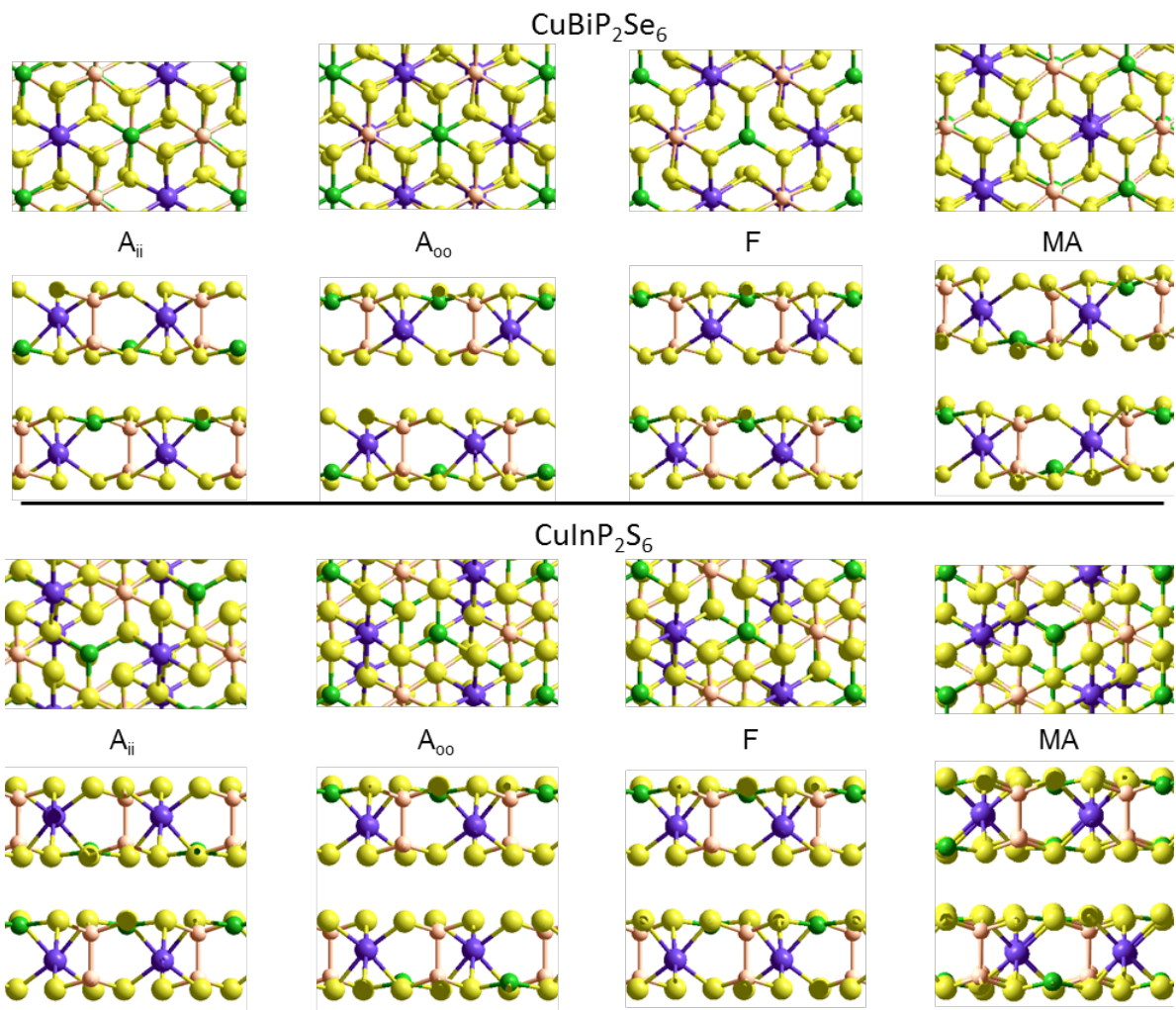

**FIG. 4**. (Color online) Top and side views of possible structures considered for bilayers of $CuBiP_2Se_6$ and $CuInP_2S_6$, (PBE-D3 optimized structures shown); green- Cu, magenta- Bi or In, orange- P, yellow- S or Se.

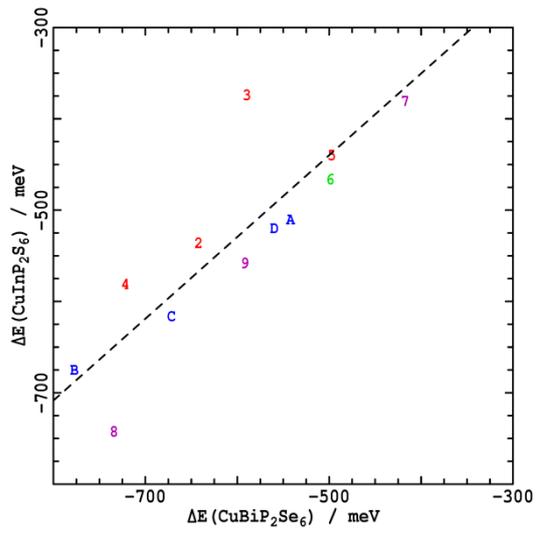

**FIG. 5.** (Color online) Comparison of calculated average bilayer interaction energies $\Delta E$ per cell for $CuInP_2S_6$ versus $CuBiP_2Se_6$, for the computational methods 0-D defined in Table I.

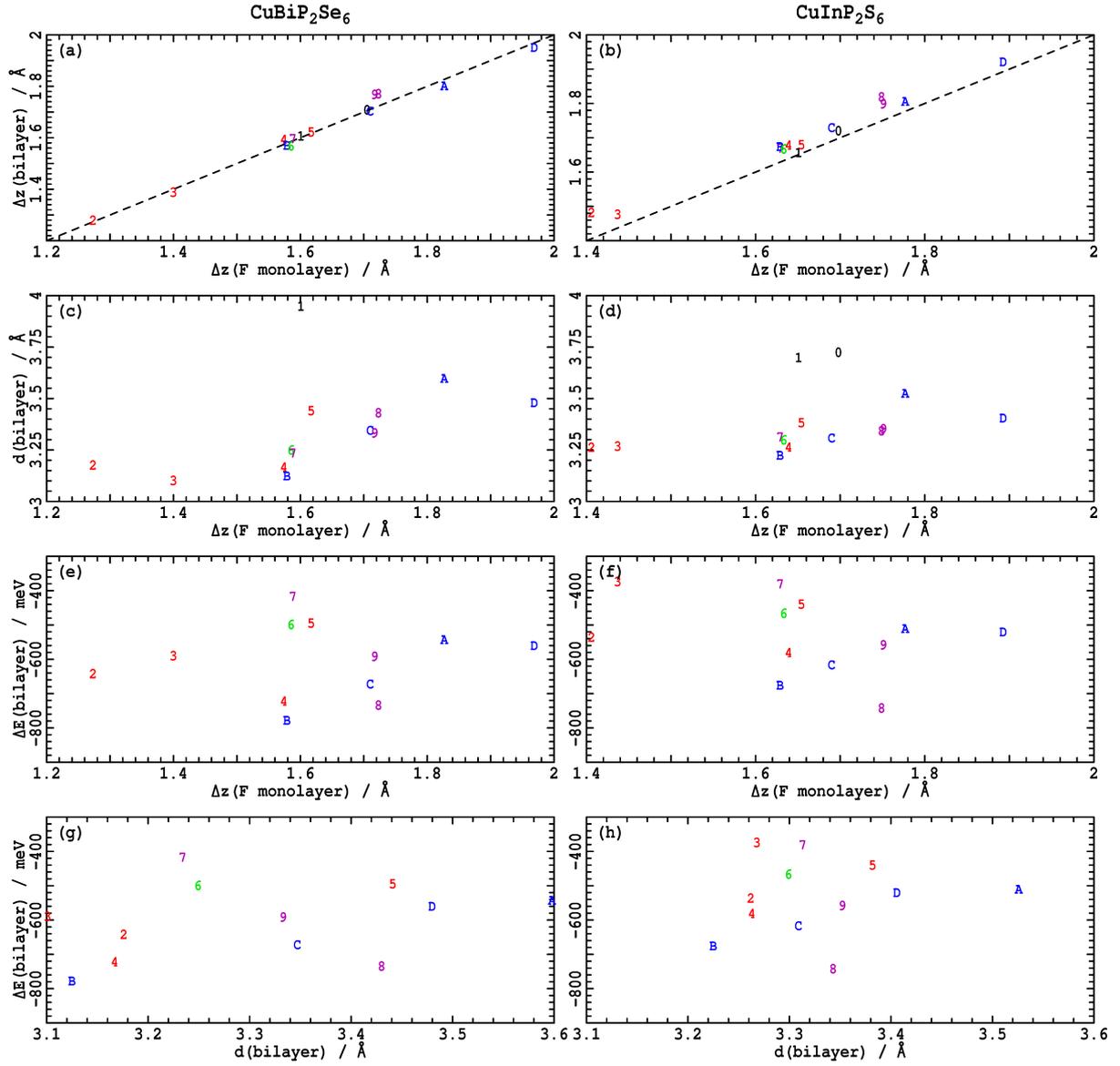

**FIG. 6.** (Color online) Correlations between monolayer and/or bilayer properties of $CuBiP_2Se_6$ and $CuInP_2S_6$, for the computational methods 0-D defined in Table I: $\Delta z$- difference in out-of-plane displacement of Cu and In/Bi atoms; $\Delta z_{CuCu}$- difference in out-of-plane displacement of the two Cu atoms in monolayer antiferroelectric $1\times\sqrt{3}$ double cells; $d$- bilayer interlayer spacing; and $\Delta E$- bilayer interaction energy between cells.

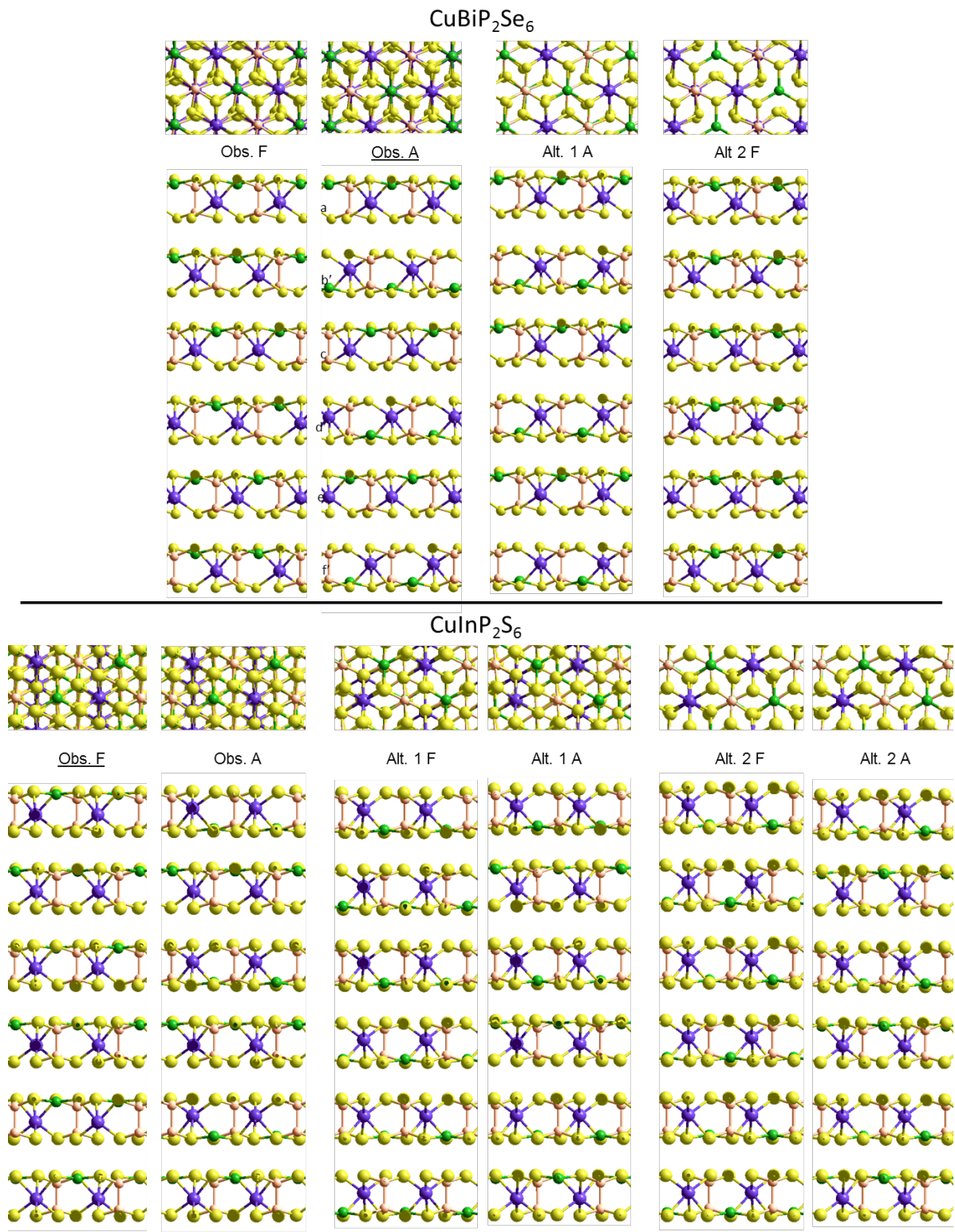

**FIG. 7.** (Color online) Top and side views of PBE-D3 optimized structures for $CuBiP_2Se_6$ and $CuInP_2S_6$ crystals, showing the conserved periodicity in the *z* direction after 6 layers. green- Cu, magenta- Bi or In, orange- P, yellow- S or Se.